# DATA SHARING, DISTRIBUTION AND UPDATING USING SOCIAL CODING COMMUNITY GITHUB AND LATEX PACKAGES IN GRADUATE RESEARCH

**Lemenkova P.**

PhD student at Charles University in Prague, Faculty of Science, Institute for Environmental Studies (Univerzita Karlova v Praze, Prírodovedecká fakulta). Benátská 2, 128 43 Praha 2, Czech Republic. E-mail: pauline.lemenkova@gmail.com

Summary. The existing code-based program implemented in GitHub portal provides a great tool for scientists and students for data sharing and notification of the co-workers, tutors and supervisors involved in research about actual updates. It enables to connect collaborators to share around current results, release datasets and updates and many more. Using standard command-line interface GitHub allows registered users to push repositories on the site. The availability of both public and private repositories enables to share current data updates with target audience: e.g., unpublished research work only for co-authors or supervisors, or, vice versa, successfully defended thesis is open for public. However, despite the evident usefulness and perspectives of GitHub, the existing users of GitHub mostly include the programmer communities and IT specialists.

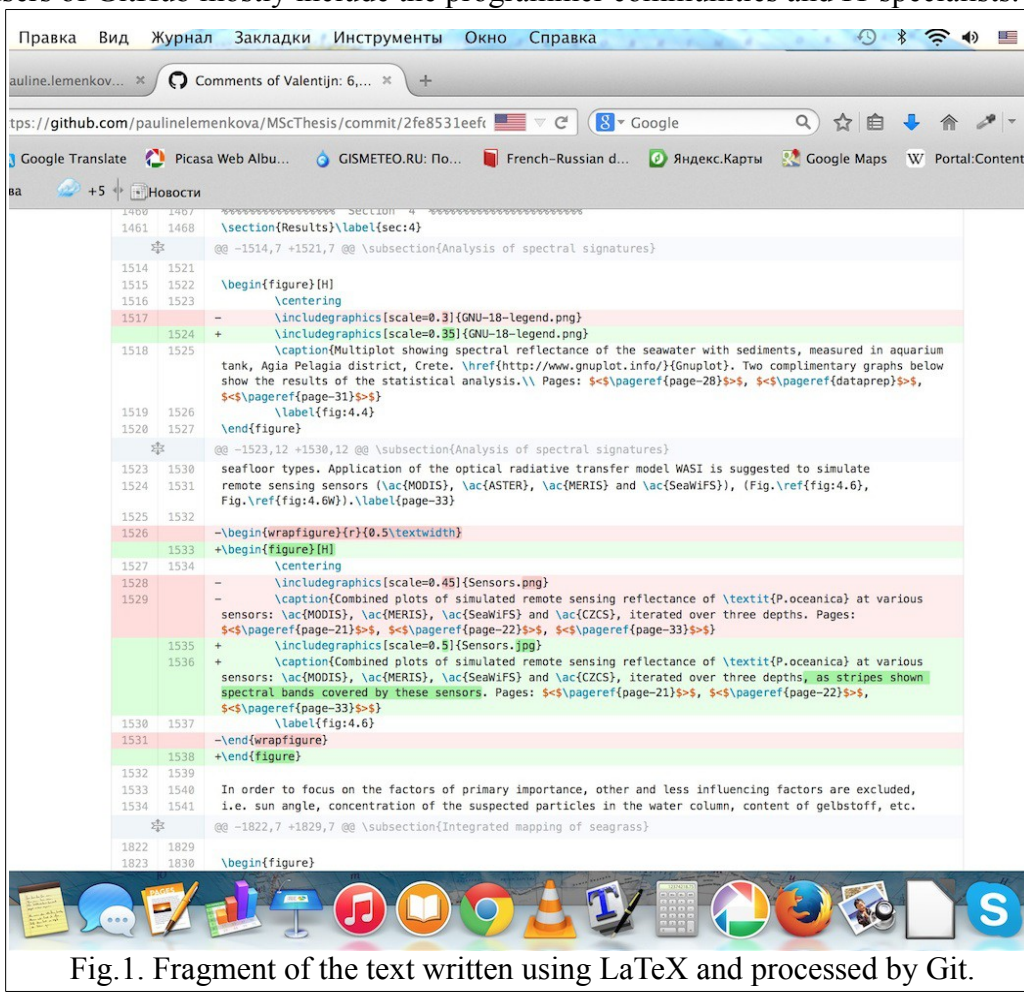

Fig.1. Fragment of the text written using LaTeX and processed by Git.

Therefore, there is a need in academic centers and universities to strongly popularize and increase the use of GitHub for student works. The case study is given on the graduate study: an MSc work successfully written and maintained using open source GitHub service at the University of Twente, Faculty of Geo-Information Science and Earth Observation (Netherlands) entitled "Seagrass monitoring and mapping along the coasts of Greece, Crete". Current presentation reports my own experience of management and organization of MSc thesis project. In spite of traditional and highly ineffective tool of MS Word, I used the effective combination of LaTeX tools with GitHub for data sharing with my tutors (Fig.1). The brilliant feature of GitHub is that it enables to indicate the latest code update since the last release (the new or modified lines of LaTeX code of my thesis were highlighted and colored, the deleted ones were indicates as well, respectively). This makes easy and effective for supervisors to take a quick look at the research progress (e.g. what has exactly been done and written since last week), to compare and revise versions. Using distributed revision control and source code management the thesis has been systematically pushed with a week-regularity for sharing with two supervisors. Flexibility and usability of open-source LaTeX packages that let user create individual projects and use various features to individualize project in combination with free account of GitHub enable student to create works. The most evident problem up-to-date consists in its non-popularity. Thus, among my classmates nobody used neither of services due to the complete ignorance or fears about its high difficulty, 'too hard to learn'. Inert, passive and stagnant idea of using MSWord is unfortunately common for many students. Using GitHub and LaTeX however gives broad possibilities for creating really interesting research and regular sharing data with target groups for participation and collaboration. This paper gives both general and technical insights regarding the use of both services highly effective for research.

1. Data sharing. The most significant feature of the portal Github lies in its possibility to share latest changes in a project online. The online repositories and web interface of Git give brilliant possibilities for supervisors, teachers, undergraduate and graduate students and researchers to write papers in a co-authorship, to store and share research papers, share and get links to recent updates of their MSc and BSc thesis, PhD dissertation, research articles and presentations. The sharing source code by email and thumb drives is no more necessary. Using GitHub enables to keep research team up to date with the latest changes to the written code. Moreover, thanks to Git's powerful merge features, the teacher don't have to worry about overwriting student's changes.

Fig.2. Illustration of "git diff" command: green colored lines
represent new additions while red colored ones – deleted parts of text

Furthermore, GitHub provides opportunities for sharing multi-source data (pictures, maps, tables, plots, graphs etc). GitHub allows co-authors to work on the current versions of the project together, to see recent changes 'on-the-fly', make changes, add comments, re-share project with new co-authors and staff. It furthermore enables both viewing and editing texts, add illustrations and different types of data, take a look at changes in the current versions, etc.

2. Control management. The use of command line, which is a standard feature in programming interface allows registered users to add current changes in a work repository of the current project. The possibility to support both public and private repository (with restricted access for welcomed audience only) provides flexibility of project management. Thus, it enables safe storage of research works that have not been published yet. At the same time it allows access to the target audience and users, which is especially important, for instance, in case of "supervisor – student" co-operation. On the contrary, successfully defended PhD dissertations, MSc and BSc theses as well as student fieldwork projects can be uploaded with access for general public and for the university archives as well. The supervisor gets links to the latest version of the student's work online and has immediate access to his work. It lets him analyze recent changes in both textual parts and graphics: new inserted data, included paragraphs, etc. Vice versa, he can control deleted and replaced parts of the text.

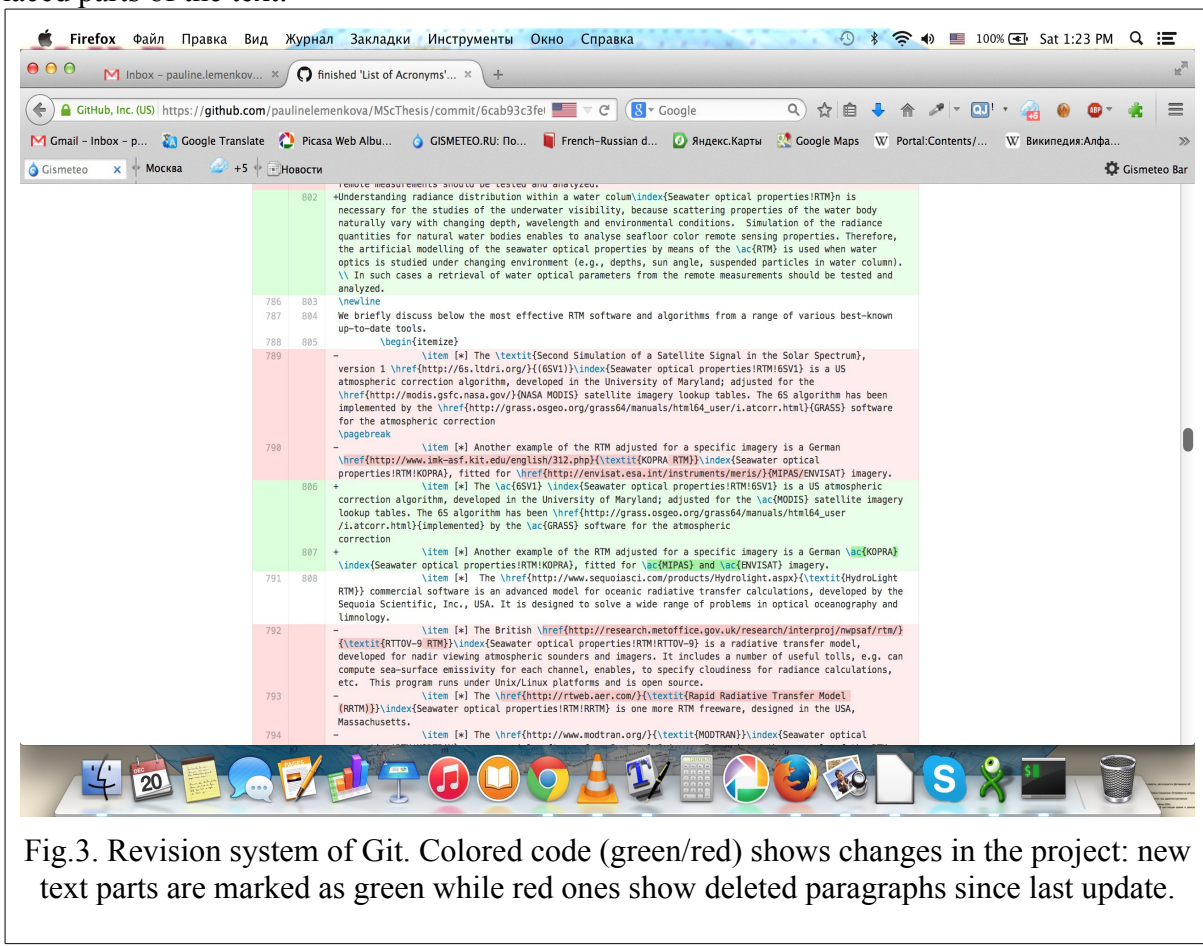

Fig.3. Revision system of Git. Colored code (green/red) shows changes in the project: new text parts are marked as green while red ones show deleted paragraphs since last update.

3. Git vs Word. There are multiple and very important drawbacks that student has to face up while using traditional software (MS Word) for writing such long research work with complex hierarchical structure and multiple-level sub-paragraphs as MSc thesis. For example, the continuous numbering and references to the illustrations will change always when user adds new ones in the middle of the text. Such normal and necessary process which always comes up during numerous re-writing and correcting of a manuscript manually again and again becomes a torture. The absence of a flexible system of cross-referencing in the bibliographic reference list of literature is a problem for the MS Word but not for LaTeX. Thus, LaTeX gives you automatic link to the specific author with correct references in the text. What's more, it updates all the links later on, even changed in the middle of the work. In case of MS Word editor, student has to check up all references to the literature again and again, which leads to a purely mechanical corrections and long, tedious and monotonous re-proofs. And what do we have in LaTeX? It has built-in embedded package BibTeX which allows to create a list of literature 'on-the-fly' and get the link to the authors instantly while citing them. The citations are editable if user needs to do some changes. Hence, the authors whose works have been used later and who were added to the list of references can later be easily added to the general list in the correct place. The citations and numeration both of previous and recent authors will be automatically modified. This is just one of many examples to illustrate advantages of using LaTeX in the educational process when writing a thesis.

4. Example of functionality GitHub and its use is detailed in this work which is based on the actual use of this service while preparing MSc thesis of the author. The thesis is written using text editor LaTeX with active use of open service Github. The combination of both services enabled regular access of supervisors to the work for monitoring progress and controlling recent changes in the text. Current paper describes personal experience in the organization, support and management of the Git repository from scratch. Unlike traditional and non-effective software (MS Word), the combined usage of LaTeX with GitHub allowed supervisors to check up the work progress on-line as soon as updates were available, to take a look at changes, insert comments and corrections.

5. LaTeX is an open source code made as a group of add-ons (packages) free for public. Its functional flexibility and convenience of use allows student to create personal creative projects or research papers using a variety of additional functions and modules. The hierarchical writing style with numerous tabulations, spaces, use of parenthesis, matching braces, indentations of subsequently written paragraphs and sub-paragraphs makes it easy for the supervisor to search through the text, find necessary parts of the text, environment of figures and tables (frames), chapters etc. As a result, consistency of such text is easier to read, analyze and go through. Excellent opportunities of Git is provided by a built-in color management code written anew. Thus, all the latest updates are syntactically colored green on the command line of Github: added text, selected lines or words, the whole fragments of paragraphs, sections and sub-sections (Fig.2). On the contrary, the red-colored sections indicate deleted text removed from the last update. Sure, it significantly helps supervisor to get a quickly assess to the work, to analyze student's progress, to ensure timely response of the student on his comments and corrections, as well as to examine what exactly has been done by the student for a given time period (e.g., in a week). The possibility of retroactive and comparative text editing system embedded in Github enables come back to the previous versions of the work, even when saved and stored a while ago. For example, if a student and a teacher want to restore some parts of the text (paragraphs, lines or even chapters), which was removed previously for some reasons or lost, it is quite possible using Git. All project versions and respectively, all changes in these versions are archived, stored and made available by Git (Fig.3). Definitely, it gives no need to worry about losing data. The latest and greatest code is always available for student and his supervisors via GitHub. Searching through the project history gives revert to a project saved, let's say, a month ago, if necessary. In this way, using the monitoring, management and revision system of Git, the master's thesis has been regularly updated and discussed with supervisors, who had regular access to project changes (Fig.4).

6. Current problems in use of GitHub. Despite evident advantages and fantastic prospects of GitHub, the majority of its users is mainly focused on professional programmers and IT-specialists. Hence, there is a need to promote Git and to demonstrate advantages of its service and environment specifically for educational and academic public which includes research and scientific labs, universities, academies, institutes and other centers of higher education, especially in the departments of computer technology, natural sciences and IT specialties. The most obvious problem in use of GitHub together with a text editor LaTeX is their unpopularity caused by following reasons. First, many students are not well informed about the existence of such services and programs and, as a result, prefer to work using traditional processor MSWord. Secondly, there is a fairly common belief about the complexity, difficulties and 'hard-to-learn' of Git and LaTeX. Though in no doubts exaggerated, the basic idea claims correct that you will do your best effort and abilities to master both of them. To learn well how to work with Git and LaTeX will definitely take certain time. However, brilliant features and advantages of using both services is obvious. It will certainly reward user for his efforts done while learning. The use of Git and LaTeX together gives best opportunities for students and their supervisors to work together in teaching-learning process: it enables regular monitoring of updates, revision control and data sharing.

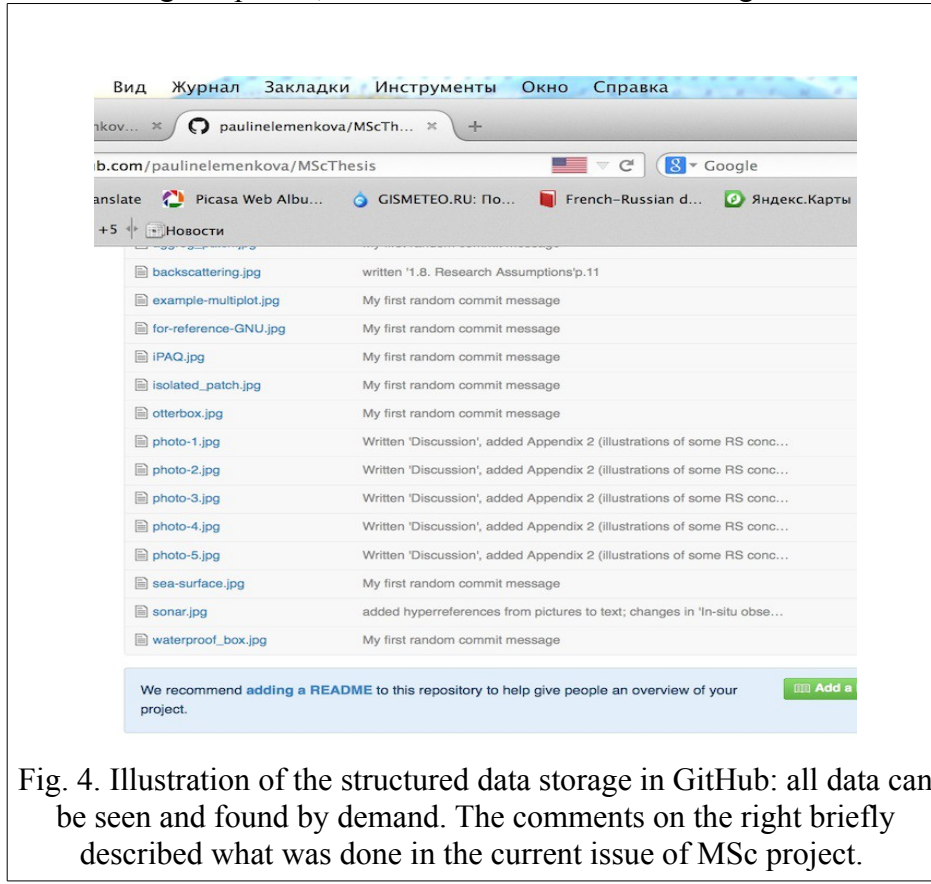

Fig. 4. Illustration of the structured data storage in GitHub: all data can be seen and found by demand. The comments on the right briefly described what was done in the current issue of MSc project.

7. Conclusion. The actuality of this work is summarized in the following outcomes. First, the paper reveals and discusses general principles of using GitHub and LaTeX in universities, advantages of their usage in education, easy access of supervisor to the student's work which helps successful supervision, controlling, monitoring and joint project running. Second, it details technical illustration of some process issues using Git and LaTeX. Third, it illustrates example of using innovative and new technologies as well as IT approach in education. The given example is illustrated by the research in the Netherlands: a case study of a successfully defended MSc work.